\documentclass[11pt,a4paper,oneside]{article}
\usepackage{graphicx,sectsty,amssymb,amsmath,color}
\usepackage[linkcolor={blue},citecolor={red},colorlinks=true]{hyperref}
\usepackage{enumerate}
\usepackage[margin=2cm]{geometry}
\usepackage{multirow}
\usepackage{authblk}
\usepackage{slashed}
\usepackage[utf8]{inputenc}
\usepackage{cite}

\definecolor{nicegreen}{rgb}{0.1,0.5,0.1}

\newcommand{\eg}{\textit{e.g.}}
\newcommand{\beq}{\begin{equation}}
\newcommand{\eeq}{\end{equation}}

\newcommand{\cO}{\mathcal{O}}
\newcommand{\cL}{\mathcal{L}}

\newcommand{\cH}{\mathcal{H}}
\newcommand{\cA}{\mathcal{A}}

\newcommand{\TeV}{\mathrm{TeV}}
\newcommand{\GeV}{\mathrm{GeV}}
\newcommand{\MeV}{\mathrm{MeV}}
\newcommand{\keV}{\mathrm{keV}}
\newcommand{\eV}{\mathrm{eV}}

\newcommand{\fm}{\mathrm{fm}}
\newcommand{\cm}{\mathrm{cm}}

\definecolor{red1}{cmyk}{0,1,1,0.3}
\definecolor{green1}{rgb}{0.3,0.8,0.3}

\newcommand{\GF}{Gamma Factory}
\newcommand{\gammaGF}{\gamma_{\rm GF}}
\newcommand{\gammaFP}{{\gamma_{\rm FP}}}
\newcommand{\Elaser}{E_{\text{\tiny laser}}}

\newcounter{mycount}
\newcommand{\myben}{\setcounter{mycount}{1}\begin{itemize}}

\newcommand{\myeen}{\end{itemize}}
\newcommand{\myconten}{\begin{itemize}}

\DeclareRobustCommand{\Fig}[1]{Fig.~\ref{#1}}

\DeclareRobustCommand{\Eq}[1]{Eq.~(\ref{#1})}

\begin{document}

\title{Probing ALPs at the CERN \GF}

\author[1]{\small Reuven Balkin}
\author[2,3]{\small Mieczyslaw W. Krasny} 
\author[1]{\small Teng Ma}
\author[4,5]{\small  Benjamin R. Safdi}
\author[1]{\small Yotam Soreq}

\affil[1]{\it\small Physics Department, Technion -- Israel Institute of Technology, Haifa 3200003, Israel}
\affil[2]{\it\small LPNHE, Sorbonne University, CNRS/IN2P3, Tour 33, RdC, 4, pl. Jussieu, 75005 Paris, France}
\affil[3]{\it\small CERN, Esplanade des Particules 1, 1211 Geneva 23, Switzerland}
\affil[4]{\it\small Berkeley Center for Theoretical Physics, University of California, Berkeley, CA 94720, U.S.A.}
\affil[5]{\it\small Theoretical Physics Group, Lawrence Berkeley National Laboratory, Berkeley, CA 94720, U.S.A.}

\maketitle

\abstract{
The aim of the proposed CERN \GF{} is to produce $\sim10^{17}$ photons per second with energies up to $400\,\MeV$. 
The photon beam intensity is expected to be a factor of $\cO(10^7)$ larger than that of  the presently available photon beams in the MeV energy range. 
In this work, we explore its potential to probe physics beyond the Standard Model. 
In particular, we discuss searches for axion like particles~(ALPs) with dominant couplings to photons and consider various production scenarios --- fixed target, photon-photon collision, and conversion by a magnetic field --- and detection schemes --- via decay to photons or back-conversion.
We find that the \GF{} in a fixed target mode can probe ALPs with mass $m_a \lesssim\cO(100\,\MeV)$ and decay constants larger than $10^7\,\GeV$, improving by an order of magnitude the discovery potential of previous beam dump experiments.
}

\section{Introduction} \label{sec:intro}

Physics beyond the Standard Model (BSM) is well motivated both by experimental evidence and by theoretical arguments;  see~\cite{Strategy:2019vxc} for recent discussions. 
One of the target goals of the \GF{} initiative~\cite{Krasny:2015ffb,GF-PoP-LoI:2019} is to produce photon beams, 
in the energy range $\cO(1-400)\,\MeV$, by colliding  partially stripped ion beams stored in the LHC with laser pulses.
If the HE-LHC project~\cite{Zimmermann:2017bbr} is realized, this energy range can be extended in the future to 1.6\,GeV. 
The \GF{} can provide a leap both in the photon beam intensity (by up to 7 orders of magnitude) and energy (by up to two orders of magnitude) with respect to the existing photon sources. 
The large photon flux in the \GF{} provides a unique opportunity to search for new particles with extremely weak couplings to the photon.
Here we propose to use the \GF{} for probing weakly  coupled,  light  pseudoscalars,  which  are  collectively  referred  to  as  axions  or  
axion-like-particles  (ALPs).
For a recent study of dark photons at the \GF{} see~\cite{Chakraborti:2021hfm}. 

ALPs are found in many well-motivated BSM models. 
One notable type of ALP is the so-called QCD axion, which was predicted as part of the Peccei-Quinn solution for the strong CP problem~\cite{Peccei:1977hh,Peccei:1977ur,Weinberg:1977ma,Wilczek:1977pj}. ALPs also appear in solutions to various hierarchy problems in the Standard Model~\cite{Graham:2015cka,Gupta:2015uea,Hook:2016mqo,Davidi:2018sii,Banerjee:2018xmn}. In some cases, ALPs can be viable dark matter candidates~\cite{Preskill:1982cy,Abbott:1982af,Dine:1982ah} or act as portals to dark sectors~\cite{Nomura:2008ru,Freytsis:2010ne,Dolan:2014ska,Hochberg:2018rjs}. 
For relevant reviews see~\cite{Marsh:2015xka,Graham:2015ouw,Hook:2018dlk,Irastorza:2018dyq,Agrawal:2021dbo}, while for recent studies of ALPs with $\sim$MeV--GeV masses, see~\cite{Dolan:2017osp,Alves:2017avw,Marciano:2016yhf,Jaeckel:2015jla,Dobrich:2015jyk,Izaguirre:2016dfi,Knapen:2016moh,Artamonov:2009sz,Fukuda:2015ana,Bauer:2018uxu,Mariotti:2017vtv,CidVidal:2018blh,Aloni:2018vki,Aloni:2019ruo,Bauer:2020jbp,Bauer:2021wjo,Sakaki:2020mqb,Florez:2021zoo,Brdar:2020dpr}.

ALPs are commonly realized as pseudo Nambu-Goldstone bosons (pNGB). 
Thus, the mass of the ALP, $m_a$, is generated as a result of a \emph{small} explicit breaking of a global symmetry, which is also spontaneously broken at some UV scale $\Lambda$, with $\Lambda \gg m_a$. 
An additional consequence of their pNGB nature is the fact that ALPs are often pseudoscalars (i.e. odd under CP). Here, we consider an ALP $a$ which couples predominantly to the photon,
\begin{align}
	\cL_a
=\frac12 \partial_\mu a \partial^\mu a -\frac12 m_a^2 a^2+	\frac{a}{4\Lambda} F_{\mu\nu} \tilde{F}^{\mu\nu}\, .
\end{align}
In this minimal setup, the lifetime of the ALP is determined by its decay width to two photons, given by
\begin{align}
\Gamma_{a\to \gamma\gamma}=\frac{m_a^3}{64\pi\Lambda^2}\,.
\end{align}
Once the ALP is produced, its lifetime in the lab frame (which takes into account also the boost factor) is crucially important when considering possible detection schemes.  
In particular, a promptly decaying ALP (on the scale of the lab) may be identified by detecting the two photons, while a long-lived ALP must first be converted back to a photon (e.g., by the inverse of the ALP production process). 
We note that an ALP-electron coupling will allow an additional visible decay channel that can increase the detection probability by decreasing the ALP lifetime; we leave this analysis for future work.  

The goal of this paper is to provide an initial assessment of the ALP discovery potential in the \GF{} 
by studying multiple production and detection strategies.
For ALP production, we consider the following possibilities (see also Fig.~\ref{fig::production}):
\begin{enumerate}
\item \label{FT} \emph{Fixed target mode} - In this production mode, the \GF{} photons (which we denote by $\gammaGF$) are collided with a high $Z$ target, labeled $N$. The axions are then produced via a Primakoff-like process, $\gammaGF\,N\to a\, N$. 
This coherent process allows the production of axions with masses $\lesssim100\,\MeV$.
\item \label{gammaconv} \emph{$B$-field conversion mode} - conversion of $\gammaGF$'s into axions in a strong magnetic field.
\item \label{gammagammacolider} \emph{Photon-photon collider mode} - The $\gammaGF$'s are collided with either another $\gammaGF$ or a laser photon originating from a Fabry-Perot~(FP) cavity, denoted as $\gammaFP$.
\item \label{Ultrarelativ} \emph{Photon-Ultra-relativistic-ion collider mode} - collisions of $\gammaGF$ with highly boosted high $Z$ ions such as lead, $\gammaGF\, {\rm Pb}\to a \, {\rm Pb}$.
\end{enumerate}

On the detection side, we consider 
\begin{enumerate}[(a)]
\item \label{detecALPdecay} ALP decays to two photons (in a beam-dump like setup), 
\item \label{detecALPconv} ALP conversion to photons in a strong magnetic field, or 
\item \label{detecALPInvPrim} ALP conversion to photons by the inverse Primakoff process, 
\end{enumerate}
see also Fig.~\ref{fig::detection}.

\section{Fixed target mode} \label{sec:FixedTarget}

\subsection{ALP production rate} 
\label{sec:ALPProd}

\begin{figure}[!t]
\begin{center}
  	\includegraphics[width=.8\textwidth]{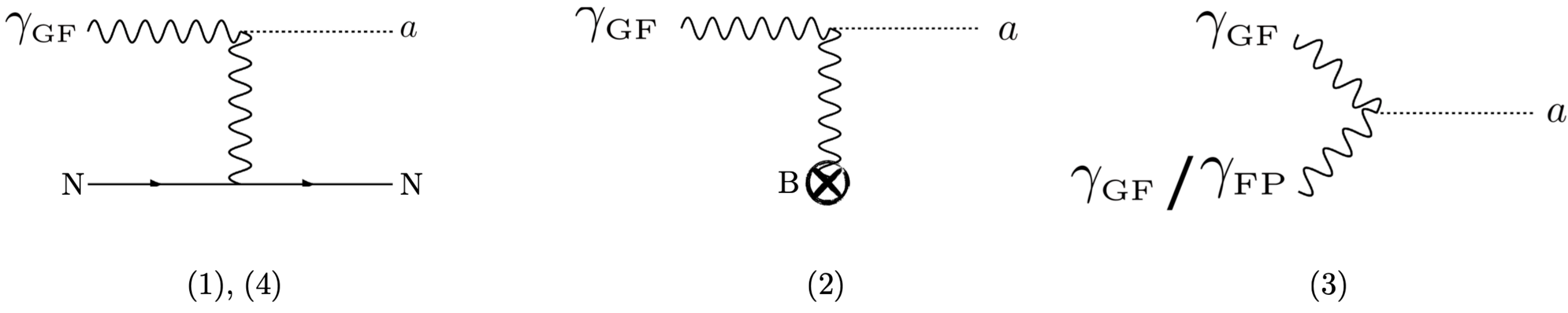} \hspace{1cm}
  	\caption{The Feynman diagrams for the different ALP production modes at the \GF{}: 
	(left)~Primakoff production relevant for fixed target~\ref{FT} and ultra-relativistic ion collision modes~\ref{Ultrarelativ}, 
	(middle)~photon conversion in a strong magnetic field~\ref{gammaconv} and
	(right)~photon-photon collider mode~\ref{gammagammacolider}. 
	 }
  	\label{fig::production}
\end{center}
\end{figure}

In this section we consider a fixed-target mode, where the \GF{} photons are collided with a fixed target, $N$.
The ALPs are produced via a Primakoff-like process
\begin{align}
	\gammaGF \, N \to a \, N\,,
\end{align}
see Fig.~\ref{fig::production}.
The elastic differential cross section is given by (\eg~in Ref.~\cite{Aloni:2019ruo})
\begin{align}
	\label{eq::XSecALPPrim}
	\frac{\mathrm{d}\sigma_{ \gamma \, N \to a \, N}}{\mathrm{d}t}
& =	\alpha Z^2 F^2(t) \Gamma_{a\to\gamma\gamma} \cH(m_N,m_a,s,t)
\end{align}
with
\begin{align}
	\cH(m_N,m_a,s,t)
=	128\pi \frac{m_N^4}{m_a^3} 
	\frac{m_a^2 t \left(m_N^2+s\right) - m_a^4 m_N^2 - t \left(\left(m_N^2-s\right)^2+s t\right)}
	{ t^2 \left(m_N^2-s\right)^2 \left(t-4 m_N^2\right)^2} \, ,
\end{align}
where $Z$ is the atomic charge, $m_N$ is the nucleus mass, $t$ and $s$ are the usual Mandelstam variables and $F(t)$ is the atomic form factor, \eg~\cite{Bjorken:2009mm,Donnelly:2017aaa,Zyla:2020zbs}. 
\begin{figure}[!t]
\begin{center}
	\includegraphics[width=.8\textwidth]{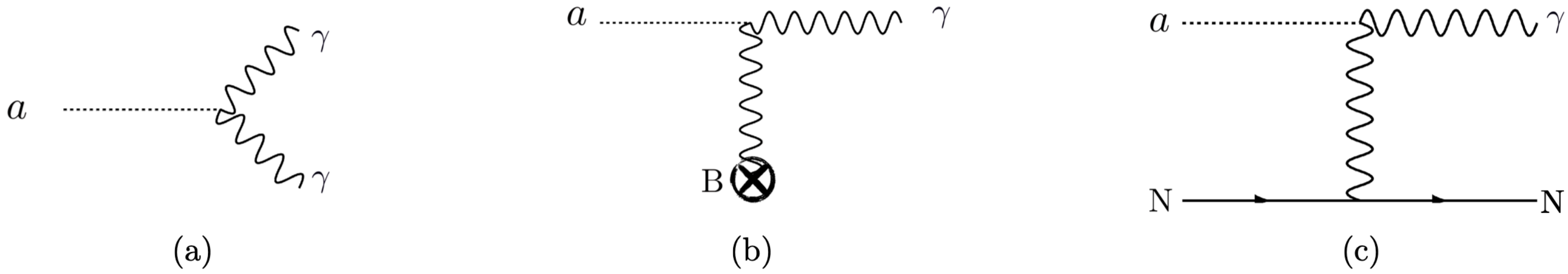} \hspace{1cm}
  	\caption{
	(left)~ALP decay into two photons~\ref{detecALPdecay},
	(middle)~ALP conversion into photon in a strong magnetic field~\ref{detecALPconv}, and
	(right)~ALP conversion into photons by the inverse Primakoff process~\ref{detecALPInvPrim}.  
	}
  	\label{fig::detection}
\end{center}
\end{figure}

The integrated luminosity of $\gammaGF -N$ collisions of $N_\gamma$ photons with a thick target material can be expressed as~\cite{Tsai:1973py}
\begin{align}
	\cL_{\rm FT}
	\approx
	N_{\gamma} \frac{\rho_N X_0}{A_N m_0}
=	19 \, {\rm nb}^{-1} \,\left( \frac{N_\gamma}{10^{24}}\right) 
	\left( \frac{\rho_N}{11.35\,{\rm g/cm^3}}\right) 
	\left( \frac{X_0}{0.56\,{\rm cm}}\right)\left( \frac{207}{A_N} \right) \, ,
\end{align}
where $m_0=1.661\times 10^{-24}\,{\rm g}$ is the nucleon mass and $\rho_N$, $A_N$ and $X_0$ are the target density, mass number and radiation length respectively. 
The benchmark values  representing the target parameters are for lead~(Pb) and were taken 
from~\cite{Zyla:2020zbs}. 
The benchmark number of photons corresponds to  the expected \GF{} rate  of $dN_\gamma/dt\sim10^{17} \, \gammaGF/\sec$, at the energy of 200\,MeV, multiplied by the effective running time of $1~\text{year} \sim 10^7\,$sec.

The total number of ALP events is given by (assuming a  monochromatic spectrum for the $\gammaGF$ for simplicity)
\begin{align}
	\label{eq:Na}
	N_a
	\approx 
	\cL_{\rm FT} \, \sigma_{\gamma \, N \to a \, N} \, \epsilon_{\rm det} \, ,
\end{align}
where $\epsilon_{\rm det}$ is the detection probability (including the angular acceptance when considering ALP decay to photons). 
Below we consider two detection schemes.
First, via ALP decay into two photons, $a\to\gamma\gamma$, see Section~(\ref{sec:ALPdecay});
second, via inverse Primakoff process, $a N \to \gamma N\,$, see Section~(\ref{sec:InvPrim}).

\subsection{Detection via ALP decays, $a \to \gamma \gamma$} 
\label{sec:ALPdecay}

In this detection scheme the target is followed by a shield (which can be of the same material or other), such that the total target and shield length is $L_S=20\,$m.
This shield is followed by the ALP decay volume of radius $R_D=2\,$m 
and length $L_D=20\,$m. 
The photon detector of radius $R_D=2\,$m is located just behind the decay volume.
The detection probability can be estimated as
\begin{align}
	\label{eq:epsagammagamma}
	\epsilon_{a\to\gamma\gamma} 
=& 	\left( e^{-L_S/L_a} - e^{-(L_D+L_S)/L_a} \right) \cA(m_a, \Lambda, p_a) \nonumber\\
	\approx&
 	2.5 \times 10^{-7} \left(\frac{L_D}{20\,{\rm m}}\right) 
	\left(\frac{0.2\,\GeV}{p_a}\right)
	\left( \frac{m_a}{10\,\MeV} \right)^4
 	\left( \frac{10^7 \,\GeV}{\Lambda} \right)^2
	\cA(m_a, \Lambda, p_a) \, ,
\end{align}
where $L_a =  p_a/(m_a\Gamma_{a\to\gamma\gamma})$ is the flight distance of the ALP in the lab frame and  $p_a\approx \sqrt{E_{\gammaGF}^2-m_a^2}$ is the ALP momentum in the lab frame (in the collinear approximation).
$\cA$ is the angular acceptance of having the two final photons inside the radius of $R_D$. 
In the second line of Eq.~\eqref{eq:epsagammagamma} we use  $L_a \gg L_D+L_S$, which is relevant to our benchmarks below. 
A shielding of 20\,m is expected to block all \GF{} photons as well as background from secondary production of SM particles inside the nuclei target. 
Thus, we consider a background free search, and the estimated projections are evaluated by requiring $N_a=3$. 

\begin{figure}[!t]
\begin{center}
  	\includegraphics[width=.6\textwidth]{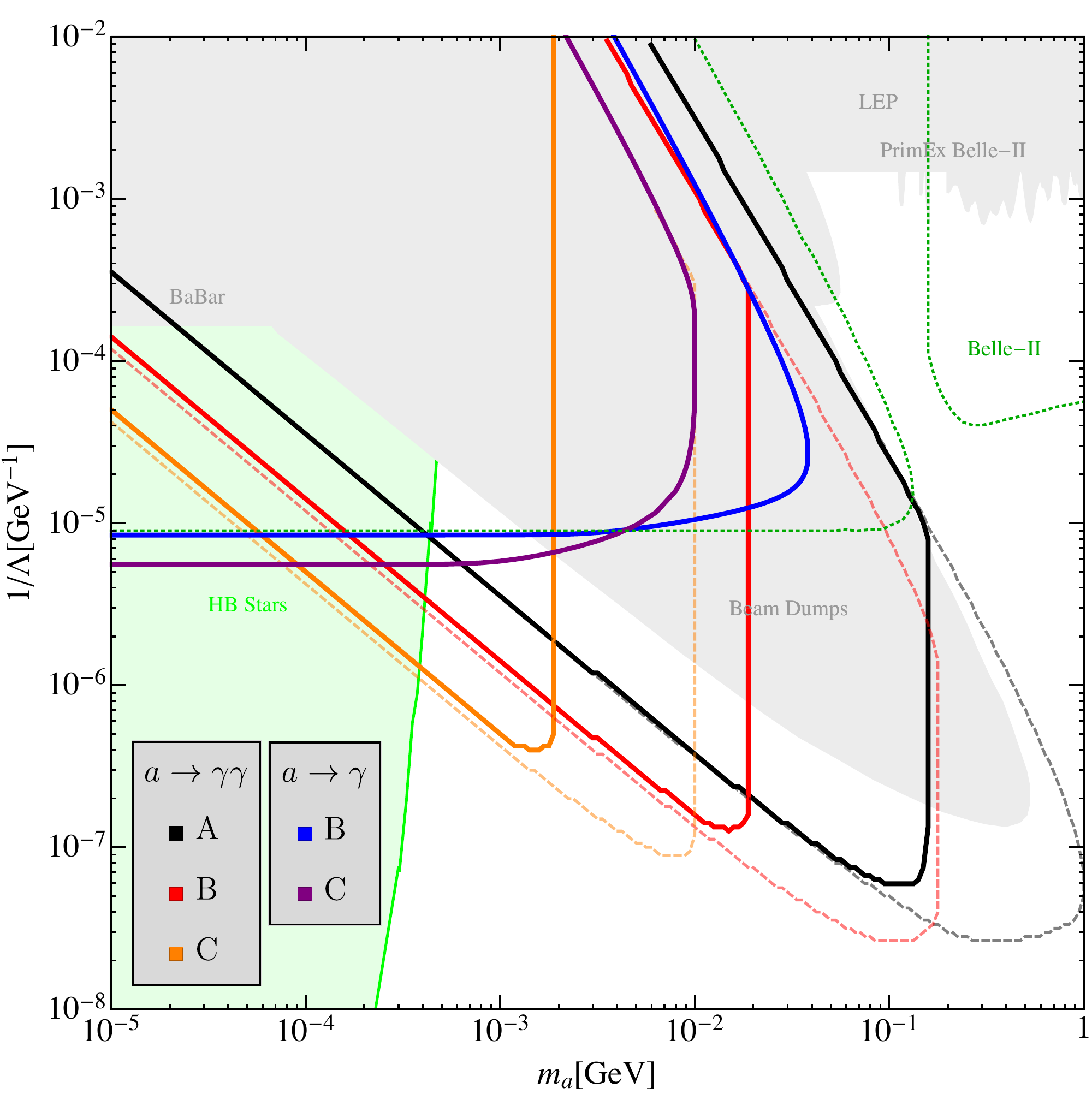} \hspace{1cm}
  	\caption{Projected sensitivity of the \GF{} in the $m_a$-$\Lambda$ plane in the fixed target production mode. 
	The black, red and orange solid curves correspond to benchmark points (A), (B) and (C), respectively, 
	in the detection scheme $a \to \gamma \gamma$ considered in Section~\ref{sec:ALPdecay}. 
	The corresponding dashed lines are the projected sensitivities in the limit of ideal angular acceptance 
	$\mathcal{A} \to 1$. 
	The blue and purple curves correspond to benchmark points (B) and (C), respectively, 
	in the detection scheme $a \to \gamma $ considered in Section~\ref{sec:InvPrim}. 
	The benchmark points are defined in Eq.~\eqref{eq:Benchmarks}. 
	The grey regions are current laboratory experimental bounds from LEP~\cite{Jaeckel:2015jla,Knapen:2016moh,Abbiendi:2002je}, {\sc PrimEx}~\cite{Larin:2010kq,Aloni:2019ruo}, NA64~\cite{Dusaev:2020gxi,Banerjee:2020fue}, Belle-II~\cite{BelleII:2020fag}, BaBar~\cite{Dolan:2017osp,Aubert:2008as} (invisible), and beam-dumps experiments~\cite{Bjorken:1988as,Blumlein:1990ay,Blumlein:1991xh}. 
	The light green region is the astrophysical bound from stellar cooling of HB stars~\cite{Carenza:2020zil}. The dark green line is the projected sensitivity of Belle-2 from the process $e\bar{e} \to \gamma+$Missing energy~\cite{Dolan:2017osp}.    }
  	\label{fig:ReachFT}
\end{center}
\end{figure}

We estimate the reach in the ALP parameter space for three benchmark scenarios, varying the energy and the flux of the \GF{} photons such that  the photon beam power, limited by the RF power of the LHC cavities~\cite{Krasny:2015ffb}, is constant. 
The benchmarks are 
\begin{align}
	\label{eq:Benchmarks}
	\text{(A):}&\ E_{\gammaGF} = 1.6\, \GeV \, ,  \qquad &\frac{dN_\gamma}{dt}=10^{16}\,{\sec^{-1}} \, , \nonumber\\
	\text{(B):}&\ E_{\gammaGF} = 0.2\, \GeV \, ,  \qquad &\frac{dN_\gamma}{dt}=10^{17}\,{\sec^{-1}} \, , \\
	\text{(C):}&\ E_{\gammaGF} = 0.02\, \GeV \, ,  \qquad &\frac{dN_\gamma}{dt}=10^{18}\,{\sec^{-1}} \,  \nonumber, 		
\end{align}
where the effective running time is assumed to be $10^7\,\sec$.
In Fig.~\ref{fig:ReachFT}, we show the projected sensitivity of the \GF{} in the $m_a$-$\Lambda$ plane.
We compare to the current bounds from laboratory experiments such as LEP~\cite{Jaeckel:2015jla,Knapen:2016moh,Abbiendi:2002je}, {\sc PrimEx}~\cite{Larin:2010kq,Aloni:2019ruo}, NA64~\cite{Dusaev:2020gxi,Banerjee:2020fue}, Belle-II~\cite{BelleII:2020fag}, BaBar~\cite{Dolan:2017osp,Aubert:2008as} (invisible), and beam-dumps experiments~\cite{Bjorken:1988as,Blumlein:1990ay,Blumlein:1991xh}. 
We also plot the astrophysical bounds from stellar cooling of horizontal branch (HB) stars~\cite{Carenza:2020zil}.
Note we chose to exclude the supernova cooling bound from SN\,1987~\cite{Raffelt:1990yz}, as its robustness is still under debate~\cite{Bar:2019ifz}.
As evident from Fig.~\ref{fig:ReachFT}, the \GF{} has the potential to probe previously unexplored ALP parameter space, \eg{} it can probe decay constants larger by a factor of up to $\cO(10)$ compared to previous beam dump experiments such as E\,137~\cite{Bjorken:1988as}. 
%

\subsection{Detection via inverse Primakoff process $  a \, N  \to \gamma\, N$} 
\label{sec:InvPrim}

The second scheme for ALP detection is via the inverse Primakoff process, $  a \, N  \to \gamma\, N$. 
This configuration resembles the so-called light-shining-thorough-wall~(LSW) setup, where the photons are essentially converted to ALPs (via the Primakoff process in this case) and converted back to photons (via the inverse process). 
The strong magnetic field in the typical LSW setup is replaced in this case by a high $Z$ nucleus target. 
This setup works as follows. 
The \GF{} photons collide with a nuclei target $N$, which is much thicker than a radiation length, and are converted to ALPs. 
After a shielding of length $L_S$, there is an \emph{instrumented} target of length $\Delta$. 
This target fulfils two purposes; it is responsible for the back-conversation of the ALPs propagating inside it and it acts as a detector for the back-converted photons.
A sandwich of lead plates interleaved with scintillator plates is the simplest example of such an instrumented target.

The above ALP detection scheme is illustrated in Fig.~\ref{fig::detection}.
This detection scheme is independent of the ALP lifetime, as long as it can reach the detection region in the target and thus can probe much lighter ALPs than the first method of Section~\ref{sec:ALPdecay}, similar to ALP invisible decays for example in $B$-factories~\cite{Dolan:2017osp,Aubert:2008as}. 

The detection probability is given by
\begin{align}
	\epsilon_{\rm InvPrim}
=	e^{-L_S/L_a} \frac{\rho_N \Delta}{A_N m_0} \sigma_{aN\to\gamma N} \cA \, ,
\end{align}
where $\cA\approx1$ is the angular acceptance, and the inverse Primakoff cross section is given by 
\begin{align}
	\frac{\sigma_{aN\to\gamma N}}{dt}
=	2\alpha Z^2 F^2(t) \Gamma_{a \to \gamma \gamma} \cH^{\prime}(m_N, m_a,s,t),  
\end{align}
with 
\begin{align}
 	\cH^{\prime}(m_{N}, m_a, s, t)  
= 	128 \pi \frac{m_N^4}{m_a ^3}  
  	\frac{m_a^2 t (m_N^2 +u)  - m_a^4 m_N^2- t( (u-m_N)^2 +ut ) }{t^2 \left( (s- m_N^2 +m_a^2)^2 -4s m_a^2 \right) (t- 4m_N^2 )^2}.
\end{align}
In Fig.~\ref{fig:ReachFT}, we show the projected sensitivity of the \GF{} in the $m_a$-$\Lambda$ plane for this ALP-detection for benchmarks (B) and (C), as defined in Eq.~\eqref{eq:Benchmarks} and $\Delta\approx 4\,$m (about 720 radiation length of lead). 
As expected (and in contrast to the setup considered in the previous subsection), the projected sensitivity is mass-independent at lower masses and extends to arbitrarily small masses, where the ALP is effectively stable at the scale of the experiment and thus can hardly be detected by observing ALP decays. 
The phase space covered by this setup is comparable to the invisible search suggested in Belle-2~\cite{Dolan:2017osp}.
Contrary to searches for indirect signals of invisible particles, which cannot be associated directly to a specific model of a hidden sector, the direct observation of the conversion of the ALP particles to photons offers an unambiguous interpretation of the observed signal.  

\section{$B$-field conversion mode}

\subsection{Production}

A photon with energy $\omega$ propagating in vacuum a distance  $L$  through a perpendicular magnetic field of magnitude $B$ has the conversion probability to ALPs of~\cite{Raffelt:1987im}
\begin{align}
	P_{\gamma \to a}(\omega,B,L) 
	\approx  
	\left(\frac{2\omega B}{ m_a^2\Lambda }\right)^2 \sin^2 \left(\frac{  m_a^2 L}{4 \omega }\right)\,.
\end{align}
This results neglects the ALP width, i.e. $\Gamma_a L \ll 1$, which we justify below for our scheme. 

\subsection{Detection} 

In this scenario we consider ALP detection through the inverse process of converting the axion back to a photon via a strong magnetic field. 
The back-conversion probability is identical. 
Thus, the total number of events (assuming ideal detection efficiency) is
\begin{align}
	N_{\text{events}} 
= 	N_{\gamma} P_{\gamma \to a}(\omega,B_1,L_1)
	P_{ a \to \gamma}(\omega,B_2,L_2) 
	\approx 
	N_{\gamma}  
	\left(\frac{2\omega B}{ m_a^2\Lambda }\right)^4 \sin^4 \left(\frac{  m_a^2 L}{4 \omega }\right)\,,
\end{align}
where for simplicity we set the parameters of both production and detection systems to be identical, namely $B_1=B_2=B$ and $L_1=L_2=L$. 
The number of events depends critically on the mass of the axion. 
We define the decoupling mass $m_a^{\text{\tiny dec.}}\equiv \sqrt{\frac{4\omega}{L}}$ and note that
\begin{align}
	P_{\gamma \to a}P_{ a \to \gamma} \begin{cases}
	\approx
	 \left(\frac{  B L}{2 \Lambda }\right)^4   \;\;\;\;\; & m_a < m_a^{\text{\tiny dec.}} \\
	\leq \left(\frac{2\omega B}{ m_a^2\Lambda }\right)^4&  m_a >  m_a^{\text{\tiny dec.}}
\end{cases}\,.
\end{align}
We find that for $m_a \ll m_a^{\text{\tiny dec.}}$ the number of events is saturated and mass-independent, while for $m_a  > m_a^{\text{\tiny dec.}}$ the conversion probability starts oscillating with an amplitude which decreases like $\propto (m_a^2\Lambda)^{-4}$. 
For the system we are considering, \eg{} for benchmark point (B), the decoupling mass is given by
\begin{align}
	m_a^{\text{\tiny dec.}} 
= 	1.6\,\eV  
	\left(\frac{\omega}{0.2\,\GeV}\right)^{1/2} 
	\left(\frac{4\cdot 15~\text{m}}{L}\right)^{1/2}.
\end{align}
At masses below $m_a^{\text{\tiny dec.}}$  we can approximate the number of events by
\begin{align}
	N_{\text{\tiny events}} 
	\approx 
	1.95 \left(\frac{N_\gamma}{10^{24}}\right)
	\left(\frac{B}{8~\text{Tesla}}\right)^{4}
	\left(\frac{L}{4\cdot 15~\text{m}}\right)^{4}
	\left(\frac{10^8~\text{GeV}}{\Lambda}\right)^{4}\,.
\end{align}
Lastly, let us justify our previous assumption of $\Gamma_a L \ll 1$. 
For the edge case of $m_a \sim m_a^{\text{\tiny dec.}}$ and $\Lambda \sim 10^3~\text{GeV}$, one finds $\Gamma_a L= \frac{m_a^3 L }{64\pi \Lambda^2}  \approx  6\cdot 10^{-18}$ \,.

The \GF{} sensitivity of $\Lambda \lesssim 10^8$~GeV is an $\cO(1)$ improvement compared to existing bound from laboratory experiments such as PVLAS~\cite{Zavattini:2007ee}, OSQAR~\cite{Ballou:2015cka} and ALPS-I~\cite{Ehret:2010mh} (see~\cite{Irastorza:2018dyq} for a summary). 
Importantly, these traditional LSW experiments, which make use of relatively low energy cavity photons, have a decoupling mass of $\cO(10^{-3})$\,eV. 
Thus, the \GF{} in the LSW configuration covers a region in parameter space, namely in the mass range $10^{-3}\,\eV \lesssim m_a \lesssim 1\,\eV$, which is not covered by experiments based on terrestrial sources for photons. 
However, this region is covered by helioscope experiments like CAST~\cite{Andriamonje:2007ew}, which benefit from the abundance of high-energy photons originating from the Sun. 
Therefore, these experiments are able to probe larger values of the UV scale, i.e. $\Lambda \gtrsim 10^{10}$\,GeV, with a similar value for the decoupling mass, i.e. $\cO(1)$\,eV.

\section{Photon-photon collider mode } 
\label{sec:photon_collider}

In this section, we estimate the reach in the $m_a$-$\Lambda$ plane for a configuration in which two beams of on-shell photons are collided. 
The total cross-section for $\gamma \gamma \to a$ can be well approximated using the narrow width approximation
\begin{align}
	\sigma(s)_{\gamma\gamma \to a } 
	\approx 
	\frac{16 \pi^2 \Gamma_a}{m_a} \delta(s-m_a^2)\,.
\end{align}
The energy spectra  of the photon beams produced by the \GF{}  in collisions of partially stripped ions with the laser pulses depends upon the ion spin and the polarisation of the laser photons~\cite{Krasny:2015ffb}. 
In the following we shall assume the simplest case of unpolarised laser photons colliding with spin-0 ions. 
The energy spectrum of such a beam is flat  and extends over the range between the laser photon energy, $\Elaser$,  and the maximal energy of $E_{\text{\tiny max}} = 4 \gamma_L ^2 \Elaser$, where $\gamma_L$ is the 
Lorentz factor of the ion beams. 
The $\Elaser$ and $\gamma_L$ values can be selected within a wide range -- optimised  for  the requisite ALP mass range. 
The $s$-dependent integrated luminosity can be expressed as $L(s)=\cL_{\rm int} \times P(s)$, where $P(s)$ is the normalised $s$-dependent luminosity distribution, 
obtained by folding the energy spectra of the colliding photon beams, and $\cL_{\rm int}$ is the integrated luminosity. The total number of events is then given by integrating over all possible values of $s$, namely $N_{\text{\tiny events}}  =\cL_{\rm int} \int ds\, \sigma(s)_{\gamma\gamma \to a }  P(s)$, which can be easily evaluated using the narrow width approximation.
Below we consider two possible configurations.
\\ \\
 {\bf Broadband mode:} 
In this configuration two colliding high-energy \GF{} photon beams have the maximal photon energy of  $E_{\text{\tiny max}} = 200\,\MeV$ providing collisions of photons in the broad range of the centre of mass energies, extended up to the value of $s_\text{max}=(2E_{\text{\tiny max}})^2$. 
We estimate the total number of events by 
\begin{align}
	N^{\text{\tiny broad}}_{\text{\tiny events}} 
= 	\cL_{\rm int}\cdot\left( \frac{16 \pi^2 \Gamma_a}{m^3_a} \right)
	\frac{\log\left( \frac{s_\text{max}}{m_a^2}\right)}{\left( \frac{s_\text{max}}{m_a^2}\right)} 
	\leq 
	3.4 \left( \frac{\mathcal{L}_{\rm int}}{10^{34} \,\text{cm}^{-2}}\right) 
	\left(\frac{10^3\,\GeV}{\Lambda} \right)^2 \;\;\; (m_a \lesssim 400\,\MeV)\,,
\end{align}
where the maximum is achieved for $m_a \approx 0.61\sqrt{s_{\text{max}}}$.  
Note that we have normalized the integrated luminosity to the \GF-reachable value  $\cL \sim 10^{27}\, \cm^{-2}\sec^{-1} $.  
If all produced axions are detected then this search may be able to provide a modest improvement in sensitivity relative to current constraints of $\Lambda\sim10^3\,\GeV$ for $50 \,\, {\rm MeV} \lesssim m_a \lesssim 400 \, \, {\rm MeV}$ (see Fig.~\ref{fig:ReachFT}).   
Note that for these masses and couplings the axions will decay promptly, at distances less than the mm scale relative to the production sights, and most of the decay photons will be at angles well away from the photon beam directions, since the axions are not significantly boosted even for the lowest axion masses relevant for this search.  
The dominant backgrounds are from $\pi^0$ production and light-by-light scattering. 
The $\pi^0$ production cross-section is larger by a few orders of magnitude relative to the axion production cross-section, across the relevant axion mass range and for $\Lambda \sim 10^3$ GeV, though the $\pi^0$ decay products should reconstruct to the pion mass and thus these events should be easily vetoed. 
The light-by-light cross-section is smaller than the axion production cross-section for low axion masses, though it may be relevant for axion masses $\sim$400 MeV.  
However,  as the invariant mass of the photons from axion decay will reconstruct the axion mass, the photons from axion decay should be distinguishable from the light-by-light photons, though for high axion masses the search may not be background free.  
In addition, we note that these region of the parameter space can be probe by Belle-2~\cite{Dolan:2017osp}, \textsc{PrimEx} and \textsc{GlueX}~\cite{Larin:2010kq,Aloni:2019ruo}.
We leave a careful investigation of the projected sensitivity of this search strategy to future work. 
\\  \\
{\bf Broadband+Narrowband mode:} 
In this configuration a \GF{} photon (with the same flat energy distribution as above) collides with a high-intensity, low-energy laser photon (\eg{} from a FP cavity). 
We simplify the calculation by assuming that the energy distribution of the laser beam is approximated by a delta-function energy distribution at  $E_0 \sim 1\,\eV$.
We estimate the total number of events by
\begin{align}
	N^{\text{\tiny b+n}}_{\text{\tiny events}} 
= 	\cL_{\rm int} \cdot\left( \frac{16 \pi^2 \Gamma_a}{m^3_a} \right)
	\frac{1}{\left( \frac{s_\text{max}}{m_a^2}\right)} 
	\leq 
	2.3 \left( \frac{\mathcal{L}_{\rm int}}{10^{44} \,\text{cm}^{-2}}\right) 
	\left(\frac{2\cdot 10^8\,\GeV}{\Lambda} \right)^2 \;\;\; (m_a \lesssim 30\,\keV)\,.
	\label{eq:broad_narrow}
\end{align}
For this mode the laser photon flux can be boosted by colliding the \GF{} photon beam, directly at its production zone,
with the intense, 20\,MHz repetition rate laser pulses stacked in the Fabry-Perot cavity.  
 As a consequence, a significantly higher value of $\cL \sim 10^{40} \cm^{-2}\sec^{-1}$  
can be delivered by the \GF{} in this collision mode. 
However, we emphasize that Eq.~\eqref{eq:broad_narrow} only provides an estimate of the number of produced axions -- the axions still must decay or be converted in order to be detected.  At these large $\Lambda \sim 10^8$ GeV, the detection efficiencies are small (see, {\it e.g.} Eq.~\eqref{eq:epsagammagamma}).  Thus, we conclude that the previously-described search strategies with the \GF~will be more sensitivite relative to this search strategy in the low mass region accessible in this collision mode.


\section{Photon-ultra-relativistic-ion collider mode } 
\label{sec:heavyion}

\begin{figure}[t!]
  \centering
    \includegraphics[width=0.5\textwidth]{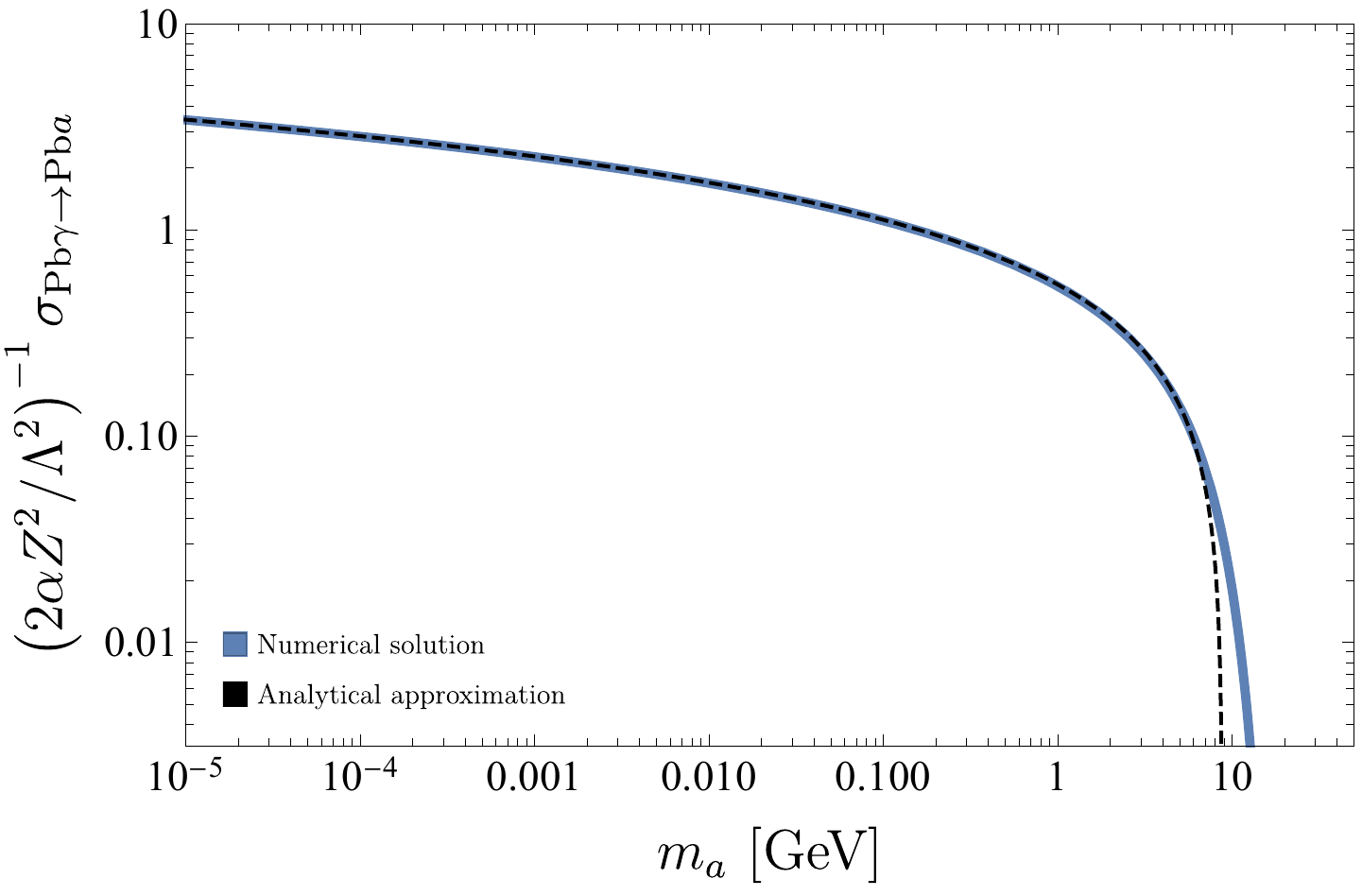}
    \caption{Total production cross section $\sigma_{ \text{Pb}\,\gamma \to  \text{Pb}\,a}$ in units of $ \frac{2\alpha Z^2 }{\Lambda^2}$. The blue curve is the numerical result of the integrated differential cross section of \Eq{eq::XSecALPPrim}. The dashed black curve is the analytical approximation for the total cross section, given in \Eq{eq::lead_axion_production_approx}.  }
    \label{fig::lead_axion_production}
\end{figure}

ALPs can be produced via a Primakoff process in which the high-energy photon collides with lead ions. 
The differential cross section is given in Eq.~\eqref{eq::XSecALPPrim} above. 
The center of mass energy for ultra-relativistic ions is given by
\begin{align}
	\label{eq::s_lead}
	\frac{s}{ m_N^2}  
	\approx  
	1+\frac{4 \,\omega A\,E_N}{m^2_N}  
= 	1+13.3 \left(\frac{\omega}{0.2\,\GeV} \right)\left( \frac{A\times E_N}{208\times 3\,\TeV}\right)
	\left(\frac{193.7\,\GeV}{m_N}\right)^2\,, 
\end{align}
where $A$ is the atomic mass number, $E_N$ is the energy per nucleon in the lab frame and $\omega$ is the energy of the photon in the lab frame. 
The ALP production is kinematically allowed when $s \geq (m_N+m_a)^2$. 
Naively, this implies an upper limit on the produced ALP mass of $m_a \lesssim  2\sqrt{\omega\, (AE_N)}-m_N \approx  540\,\GeV$, where we use the parameters in \Eq{eq::s_lead}.
However, the process is effectively cutoff at large momenta exchange due to the atomic factor $F(t)$, which leads to a lower upper limit on the produced ALP mass. 
At the qualitative level, we can treat the form factor as a step function which cuts off when the transferred momentum is at the order of the inverse of the nucleus size, i.e. $q\sim\cO(\fm^{-1})$.
Thus, for lead we have
\begin{align}
	\label{eq::tmax_coherence}
	|t| \lesssim - \bar{t}_{\text{\tiny max}} \sim (0.03\,\GeV)^2\,,
\end{align}
where we use the charge density parameterization and values from~\cite{Jones:2014aoa}. 
This upper limit implies an upper limit on the produced ALP mass. The maximal ALP mass can be calculated by solving $\bar{t}_{\text{\tiny max}}=\frac{ m^4_a  }{4s} - \left( k_{\rm cm} + p_{\rm cm} \right)^2$, which is the maximal allowed $t$, where $k_{\rm cm}\,(p_{\rm cm})$ is the photon\,(ALP) momentum in the center of mass frame.  In the limit of $|t_{\text{\tiny max}}| \ll m_N^2 \ll s$, we find that
\begin{align}
	\label{eq::coherence_upper_limit}
	\overline{m}_a^2 
	\approx 
	s \sqrt{\frac{-\bar{t}_{\text{\tiny max}}}{m_N^2}}
	\approx 
	\frac{4 \omega A E_N  \sqrt{-\bar{t}_{\text{\tiny max}}}}{m_N} 
	\approx 
	(8.8\,\GeV)^2\;\;\; (\text{coherence upper limit})\,, 
\end{align}
where in the last step we used the same input parameters as in \Eq{eq::s_lead}. 
The total cross section is calculated by integrating over all scattering angles encoded in $t$. Approximating the nuclear form factor $F(t)$ as a step function which cuts integration at $\bar{t}_{\text{\tiny max}}$, the total cross section can be approximated (to leading order in $m_N^2/s$) as 
\begin{align}
	\label{eq::lead_axion_production_approx}
	\sigma_{ \text{Pb}\,\gamma \to  \text{Pb}\,a}  
	\approx  
	\left( \frac{2\alpha Z^2 }{\Lambda^2}\right) \times 
	\frac{1}{16}\log \left(\frac{\bar{t}_{\text{\tiny max}}}{{t}_{\text{\tiny min}}} \right) 
	\approx  
	\left( \frac{2\alpha Z^2 }{\Lambda^2}\right) \times  \frac14\log \left(\frac{\overline{m}_a }{m_a} \right) \,. 
\end{align}
We illustrate the total cross section as a function of the axion mass in \Fig{fig::lead_axion_production}. 
The analytical approximation describes the full numerical result up to small $\cO(1\%)$ corrections. 
The approximation only breaks down near the threshold region $m_a \sim \bar{m}_a$, where the leading $\log$ vanishes and sub-leading corrections begin to dominate. 
A similar estimation for $\overline{m}_a^2$ of \Eq{eq::coherence_upper_limit} can be done by using the equivalent photon approximation~\cite{vonWeizsacker:1934nji,Williams:1934ad}. 

In it interesting to compare our setup to the setup suggested in Ref.~\cite{Knapen:2016moh}, where ALPs are produced in collisions of the two counter-propagating ion beams.
Assuming the same energy per nucleon $E_N = 3~\text{TeV}$ (as opposed to the $5.5~\text{TeV}$ considered in Ref.~\cite{Knapen:2016moh}), the (equivalent) photons have a maximal energy of $\omega_{\text{\tiny max}} = 2\gamma/R_A$, where $\gamma={A E_N}/{m_N}$ is the boost factor of the ion in the lab frame and $R_A =  0.028~\text{GeV} (A/208)^{1/3}$ is the nuclear radius. Thus, the heaviest axion that can be produced from the effective 2-photon system originating from Pb-Pb collision is
\begin{align}
m^2_a \lesssim s_{\text{\tiny max}} =  (2\omega_{\text{\tiny max}})^2 = \left(\frac{2\gamma}{R_A}\right)^2 = \left(\frac{2 A E_N}{m_N R_A}\right)^2 \approx (180~\text{GeV})^2 \;\;\; \text{(Pb-Pb collisions)}\,.
\end{align}
Therefore, the lead-lead collision are capable of producing axions there are roughly an order of magnitude heavier.\footnote{Another notable difference is that the cross-section for the Pb-Pb collision is a factor $\alpha Z^2\sim 50$ enhanced with respect to the Pb-$\gamma$ cross-section, which can potentially increase the sensitivity to $\Lambda$ with respect to our setup. However, when comparing the total number of events, this enhancement is partially cancelled by the lower luminosity expected in Pb-Pb collision $\mathcal{L}_{\text{Pb-Pb}} \sim 1~\text{nb}^{-1}/\text{year}$~\cite{Knapen:2016moh} compared to our benchmark luminosity of $\mathcal{L}_{\gamma} \sim 10~\text{nb}^{-1}/\text{year}$.  }

Let us estimate the reach of this setup, assuming for simplicity ideal detection efficiency and zero background. The total number of axions produced is given by
\begin{align}
	N_{\text{\tiny events}} 
	\approx  
	\cL \cdot \left( \frac{2\alpha Z^2 }{\Lambda^2}\right) 
=	3.8 \left(\frac{\cL}{ 10^{34} \cm^{-2} \text{year}^{-1}} \right)\left(\frac{10^4~\text{GeV}}{\Lambda}\right)^2\,.
\end{align}
The projected sensitivity of $\Lambda \sim 10^4$~GeV  would be comparable to but stronger, potentially by around an order of magnitude, than existing experimental bounds at these axion masses (see Fig.~\ref{fig:ReachFT}).  In particular, this search strategy may have an improved reach, both in $\Lambda$ and $m_a$, relative to the photon-photon collider mode discussed in Sec.~\ref{sec:photon_collider}.  However, more work is needed to assess possible backgrounds and detection efficiencies.

\section{Outlook} 
\label{sec:outlook}

In this work we demonstrate that the \GF{} has the potential to probe unexplored regions of ALP parameter space. 
In particular, the fixed target mode is sensitive to decay constants, for ALPs with mass $m_a \lesssim\cO(100\,\MeV)$, that are up to $\cO(10)$ larger compared to those probed by existing beam dump experiments.
The heavy ion and photon-photon collider modes may also improve upon existing bounds, though the LSW mode appear less promising at present accounting for current laboratory and astrophysical constraints. 

In the above study, we focus on a minimal model where the ALP couples dominantly to photons. 
However, one can go beyond and consider coupling to leptons, quark and gluons as well. 
For example, if the ALP couples to gluons, it can be produced or detected via vector meson-photon mixing and vector meson exchange in the $t$-channel~\cite{Aloni:2019ruo}. 
We leave this and other non-minimal coupling scenarios for future work.

\section*{Acknowledgements}

We thank Dima Budker and Gilad Perez for fruitful discussions. 
RB, TM and YS  are supported by grants from NSF-BSF, ISF and the Azrieli foundation. 
TM is supported by ``Study in Israel" Fellowship for Outstanding Post-Doctoral Researchers from China and India by PBC of CHE.
BRS  was  supported  in  part  by the US DOE  Early  Career  Grant  DESC0019225. 
YS is a Taub fellow (supported by the Taub Family Foundation).

\bibliographystyle{utphys}
\bibliography{GF_BSM}

\providecommand{\href}[2]{#2}\begingroup\raggedright\begin{thebibliography}{10}

\bibitem{Strategy:2019vxc}
R.~K. Ellis {\em et~al.}, ``{Physics Briefing Book}: {Input for the European
  Strategy for Particle Physics Update 2020},''
  \href{http://arxiv.org/abs/1910.11775}{{\ttfamily arXiv:1910.11775
  [hep-ex]}}.

\bibitem{Krasny:2015ffb}
M.~W. Krasny, ``{\em The Gamma Factory proposal for CERN},'' 2015.
\newblock {arXiv:1511.07794 [hep-ex]}.

\bibitem{GF-PoP-LoI:2019}
{ Gamma Factory Study Group} Collaboration, M.~W. Krasny and {\rm et
  al.}~(Gamma Factory Study~Group), ``{\it Gamma Factory Proof-of-Principle
  experiment}.''. Letter-of-Intent (LoI), CERN-SPSC-2019-031, SPSC-I-253, {\bf
  2019}.

\bibitem{Zimmermann:2017bbr}
F.~Zimmermann, ``{HE-LHC Overview, Parameters and Challenges},'' {\em ICFA Beam
  Dyn. Newslett.} {\bfseries 72} (2017) 138--141.

\bibitem{Chakraborti:2021hfm}
S.~Chakraborti, J.~L. Feng, J.~K. Koga, and M.~Valli, ``{Gamma Factory Searches
  for Extremely Weakly-Interacting Particles},''
  \href{http://arxiv.org/abs/2105.10289}{{\ttfamily arXiv:2105.10289
  [hep-ph]}}.

\bibitem{Peccei:1977hh}
R.~D. Peccei and H.~R. Quinn, ``{CP Conservation in the Presence of
  Instantons},'' \href{http://dx.doi.org/10.1103/PhysRevLett.38.1440}{{\em
  Phys. Rev. Lett.} {\bfseries 38} (1977) 1440--1443}.

\bibitem{Peccei:1977ur}
R.~D. Peccei and H.~R. Quinn, ``{Constraints Imposed by CP Conservation in the
  Presence of Instantons},''
  \href{http://dx.doi.org/10.1103/PhysRevD.16.1791}{{\em Phys. Rev. D}
  {\bfseries 16} (1977) 1791--1797}.

\bibitem{Weinberg:1977ma}
S.~Weinberg, ``{A New Light Boson?},''
  \href{http://dx.doi.org/10.1103/PhysRevLett.40.223}{{\em Phys. Rev. Lett.}
  {\bfseries 40} (1978) 223--226}.

\bibitem{Wilczek:1977pj}
F.~Wilczek, ``{Problem of Strong $P$ and $T$ Invariance in the Presence of
  Instantons},'' \href{http://dx.doi.org/10.1103/PhysRevLett.40.279}{{\em Phys.
  Rev. Lett.} {\bfseries 40} (1978) 279--282}.

\bibitem{Graham:2015cka}
P.~W. Graham, D.~E. Kaplan, and S.~Rajendran, ``{Cosmological Relaxation of the
  Electroweak Scale},''
  \href{http://dx.doi.org/10.1103/PhysRevLett.115.221801}{{\em Phys. Rev.
  Lett.} {\bfseries 115} no.~22, (2015) 221801},
  \href{http://arxiv.org/abs/1504.07551}{{\ttfamily arXiv:1504.07551
  [hep-ph]}}.

\bibitem{Gupta:2015uea}
R.~S. Gupta, Z.~Komargodski, G.~Perez, and L.~Ubaldi, ``{Is the Relaxion an
  Axion?},'' \href{http://dx.doi.org/10.1007/JHEP02(2016)166}{{\em JHEP}
  {\bfseries 02} (2016) 166}, \href{http://arxiv.org/abs/1509.00047}{{\ttfamily
  arXiv:1509.00047 [hep-ph]}}.

\bibitem{Hook:2016mqo}
A.~Hook and G.~Marques-Tavares, ``{Relaxation from particle production},''
  \href{http://dx.doi.org/10.1007/JHEP12(2016)101}{{\em JHEP} {\bfseries 12}
  (2016) 101}, \href{http://arxiv.org/abs/1607.01786}{{\ttfamily
  arXiv:1607.01786 [hep-ph]}}.

\bibitem{Davidi:2018sii}
O.~Davidi, R.~S. Gupta, G.~Perez, D.~Redigolo, and A.~Shalit, ``{The
  hierarchion, a relaxion addressing the Standard Model\textquoteright{}s
  hierarchies},'' \href{http://dx.doi.org/10.1007/JHEP08(2018)153}{{\em JHEP}
  {\bfseries 08} (2018) 153}, \href{http://arxiv.org/abs/1806.08791}{{\ttfamily
  arXiv:1806.08791 [hep-ph]}}.

\bibitem{Banerjee:2018xmn}
A.~Banerjee, H.~Kim, and G.~Perez, ``{Coherent relaxion dark matter},''
  \href{http://dx.doi.org/10.1103/PhysRevD.100.115026}{{\em Phys. Rev. D}
  {\bfseries 100} no.~11, (2019) 115026},
  \href{http://arxiv.org/abs/1810.01889}{{\ttfamily arXiv:1810.01889
  [hep-ph]}}.

\bibitem{Preskill:1982cy}
J.~Preskill, M.~B. Wise, and F.~Wilczek, ``{Cosmology of the Invisible
  Axion},'' \href{http://dx.doi.org/10.1016/0370-2693(83)90637-8}{{\em Phys.
  Lett. B} {\bfseries 120} (1983) 127--132}.

\bibitem{Abbott:1982af}
L.~F. Abbott and P.~Sikivie, ``{A Cosmological Bound on the Invisible Axion},''
  \href{http://dx.doi.org/10.1016/0370-2693(83)90638-X}{{\em Phys. Lett. B}
  {\bfseries 120} (1983) 133--136}.

\bibitem{Dine:1982ah}
M.~Dine and W.~Fischler, ``{The Not So Harmless Axion},''
  \href{http://dx.doi.org/10.1016/0370-2693(83)90639-1}{{\em Phys. Lett. B}
  {\bfseries 120} (1983) 137--141}.

\bibitem{Nomura:2008ru}
Y.~Nomura and J.~Thaler, ``{Dark Matter through the Axion Portal},''
  \href{http://dx.doi.org/10.1103/PhysRevD.79.075008}{{\em Phys. Rev. D}
  {\bfseries 79} (2009) 075008},
  \href{http://arxiv.org/abs/0810.5397}{{\ttfamily arXiv:0810.5397 [hep-ph]}}.

\bibitem{Freytsis:2010ne}
M.~Freytsis and Z.~Ligeti, ``{On dark matter models with uniquely
  spin-dependent detection possibilities},''
  \href{http://dx.doi.org/10.1103/PhysRevD.83.115009}{{\em Phys. Rev. D}
  {\bfseries 83} (2011) 115009},
  \href{http://arxiv.org/abs/1012.5317}{{\ttfamily arXiv:1012.5317 [hep-ph]}}.

\bibitem{Dolan:2014ska}
M.~J. Dolan, F.~Kahlhoefer, C.~McCabe, and K.~Schmidt-Hoberg, ``{A taste of
  dark matter: Flavour constraints on pseudoscalar mediators},''
  \href{http://dx.doi.org/10.1007/JHEP03(2015)171}{{\em JHEP} {\bfseries 03}
  (2015) 171}, \href{http://arxiv.org/abs/1412.5174}{{\ttfamily arXiv:1412.5174
  [hep-ph]}}. [Erratum: JHEP 07, 103 (2015)].

\bibitem{Hochberg:2018rjs}
Y.~Hochberg, E.~Kuflik, R.~Mcgehee, H.~Murayama, and K.~Schutz, ``{Strongly
  interacting massive particles through the axion portal},''
  \href{http://dx.doi.org/10.1103/PhysRevD.98.115031}{{\em Phys. Rev. D}
  {\bfseries 98} no.~11, (2018) 115031},
  \href{http://arxiv.org/abs/1806.10139}{{\ttfamily arXiv:1806.10139
  [hep-ph]}}.

\bibitem{Marsh:2015xka}
D.~J.~E. Marsh, ``{Axion Cosmology},''
  \href{http://dx.doi.org/10.1016/j.physrep.2016.06.005}{{\em Phys. Rept.}
  {\bfseries 643} (2016) 1--79},
  \href{http://arxiv.org/abs/1510.07633}{{\ttfamily arXiv:1510.07633
  [astro-ph.CO]}}.

\bibitem{Graham:2015ouw}
P.~W. Graham, I.~G. Irastorza, S.~K. Lamoreaux, A.~Lindner, and K.~A. van
  Bibber, ``{Experimental Searches for the Axion and Axion-Like Particles},''
  \href{http://dx.doi.org/10.1146/annurev-nucl-102014-022120}{{\em Ann. Rev.
  Nucl. Part. Sci.} {\bfseries 65} (2015) 485--514},
  \href{http://arxiv.org/abs/1602.00039}{{\ttfamily arXiv:1602.00039
  [hep-ex]}}.

\bibitem{Hook:2018dlk}
A.~Hook, ``{TASI Lectures on the Strong CP Problem and Axions},'' {\em PoS}
  {\bfseries TASI2018} (2019) 004,
  \href{http://arxiv.org/abs/1812.02669}{{\ttfamily arXiv:1812.02669
  [hep-ph]}}.

\bibitem{Irastorza:2018dyq}
I.~G. Irastorza and J.~Redondo, ``{New experimental approaches in the search
  for axion-like particles},''
  \href{http://dx.doi.org/10.1016/j.ppnp.2018.05.003}{{\em Prog. Part. Nucl.
  Phys.} {\bfseries 102} (2018) 89--159},
  \href{http://arxiv.org/abs/1801.08127}{{\ttfamily arXiv:1801.08127
  [hep-ph]}}.

\bibitem{Agrawal:2021dbo}
P.~Agrawal {\em et~al.}, ``{Feebly-Interacting Particles:FIPs 2020 Workshop
  Report},'' \href{http://arxiv.org/abs/2102.12143}{{\ttfamily arXiv:2102.12143
  [hep-ph]}}.

\bibitem{Dolan:2017osp}
M.~J. Dolan, T.~Ferber, C.~Hearty, F.~Kahlhoefer, and K.~Schmidt-Hoberg,
  ``{Revised constraints and Belle II sensitivity for visible and invisible
  axion-like particles},''
  \href{http://dx.doi.org/10.1007/JHEP12(2017)094}{{\em JHEP} {\bfseries 12}
  (2017) 094}, \href{http://arxiv.org/abs/1709.00009}{{\ttfamily
  arXiv:1709.00009 [hep-ph]}}. [Erratum: JHEP 03, 190 (2021)].

\bibitem{Alves:2017avw}
D.~S.~M. Alves and N.~Weiner, ``{A viable QCD axion in the MeV mass range},''
  \href{http://dx.doi.org/10.1007/JHEP07(2018)092}{{\em JHEP} {\bfseries 07}
  (2018) 092}, \href{http://arxiv.org/abs/1710.03764}{{\ttfamily
  arXiv:1710.03764 [hep-ph]}}.

\bibitem{Marciano:2016yhf}
W.~J. Marciano, A.~Masiero, P.~Paradisi, and M.~Passera, ``{Contributions of
  axionlike particles to lepton dipole moments},''
  \href{http://dx.doi.org/10.1103/PhysRevD.94.115033}{{\em Phys. Rev. D}
  {\bfseries 94} no.~11, (2016) 115033},
  \href{http://arxiv.org/abs/1607.01022}{{\ttfamily arXiv:1607.01022
  [hep-ph]}}.

\bibitem{Jaeckel:2015jla}
J.~Jaeckel and M.~Spannowsky, ``{Probing MeV to 90 GeV axion-like particles
  with LEP and LHC},''
  \href{http://dx.doi.org/10.1016/j.physletb.2015.12.037}{{\em Phys. Lett. B}
  {\bfseries 753} (2016) 482--487},
  \href{http://arxiv.org/abs/1509.00476}{{\ttfamily arXiv:1509.00476
  [hep-ph]}}.

\bibitem{Dobrich:2015jyk}
B.~D\"obrich, J.~Jaeckel, F.~Kahlhoefer, A.~Ringwald, and K.~Schmidt-Hoberg,
  ``{ALPtraum: ALP production in proton beam dump experiments},''
  \href{http://dx.doi.org/10.1007/JHEP02(2016)018}{{\em JHEP} {\bfseries 02}
  (2016) 018}, \href{http://arxiv.org/abs/1512.03069}{{\ttfamily
  arXiv:1512.03069 [hep-ph]}}.

\bibitem{Izaguirre:2016dfi}
E.~Izaguirre, T.~Lin, and B.~Shuve, ``{Searching for Axionlike Particles in
  Flavor-Changing Neutral Current Processes},''
  \href{http://dx.doi.org/10.1103/PhysRevLett.118.111802}{{\em Phys. Rev.
  Lett.} {\bfseries 118} no.~11, (2017) 111802},
  \href{http://arxiv.org/abs/1611.09355}{{\ttfamily arXiv:1611.09355
  [hep-ph]}}.

\bibitem{Knapen:2016moh}
S.~Knapen, T.~Lin, H.~K. Lou, and T.~Melia, ``{Searching for Axionlike
  Particles with Ultraperipheral Heavy-Ion Collisions},''
  \href{http://dx.doi.org/10.1103/PhysRevLett.118.171801}{{\em Phys. Rev.
  Lett.} {\bfseries 118} no.~17, (2017) 171801},
  \href{http://arxiv.org/abs/1607.06083}{{\ttfamily arXiv:1607.06083
  [hep-ph]}}.

\bibitem{Artamonov:2009sz}
{ BNL-E949} Collaboration, A.~V. Artamonov {\em et~al.}, ``{Study of the decay
  $K^+\to\pi^+\nu \bar\nu$ in the momentum region $140 < P_\pi < 199$ MeV/c},''
  \href{http://dx.doi.org/10.1103/PhysRevD.79.092004}{{\em Phys. Rev. D}
  {\bfseries 79} (2009) 092004},
  \href{http://arxiv.org/abs/0903.0030}{{\ttfamily arXiv:0903.0030 [hep-ex]}}.

\bibitem{Fukuda:2015ana}
H.~Fukuda, K.~Harigaya, M.~Ibe, and T.~T. Yanagida, ``{Model of visible QCD
  axion},'' \href{http://dx.doi.org/10.1103/PhysRevD.92.015021}{{\em Phys. Rev.
  D} {\bfseries 92} no.~1, (2015) 015021},
  \href{http://arxiv.org/abs/1504.06084}{{\ttfamily arXiv:1504.06084
  [hep-ph]}}.

\bibitem{Bauer:2018uxu}
M.~Bauer, M.~Heiles, M.~Neubert, and A.~Thamm, ``{Axion-Like Particles at
  Future Colliders},''
  \href{http://dx.doi.org/10.1140/epjc/s10052-019-6587-9}{{\em Eur. Phys. J. C}
  {\bfseries 79} no.~1, (2019) 74},
  \href{http://arxiv.org/abs/1808.10323}{{\ttfamily arXiv:1808.10323
  [hep-ph]}}.

\bibitem{Mariotti:2017vtv}
A.~Mariotti, D.~Redigolo, F.~Sala, and K.~Tobioka, ``{New LHC bound on low-mass
  diphoton resonances},''
  \href{http://dx.doi.org/10.1016/j.physletb.2018.06.039}{{\em Phys. Lett. B}
  {\bfseries 783} (2018) 13--18},
  \href{http://arxiv.org/abs/1710.01743}{{\ttfamily arXiv:1710.01743
  [hep-ph]}}.

\bibitem{CidVidal:2018blh}
X.~Cid~Vidal, A.~Mariotti, D.~Redigolo, F.~Sala, and K.~Tobioka, ``{New Axion
  Searches at Flavor Factories},''
  \href{http://dx.doi.org/10.1007/JHEP01(2019)113}{{\em JHEP} {\bfseries 01}
  (2019) 113}, \href{http://arxiv.org/abs/1810.09452}{{\ttfamily
  arXiv:1810.09452 [hep-ph]}}. [Erratum: JHEP 06, 141 (2020)].

\bibitem{Aloni:2018vki}
D.~Aloni, Y.~Soreq, and M.~Williams, ``{Coupling QCD-Scale Axionlike Particles
  to Gluons},'' \href{http://dx.doi.org/10.1103/PhysRevLett.123.031803}{{\em
  Phys. Rev. Lett.} {\bfseries 123} no.~3, (2019) 031803},
  \href{http://arxiv.org/abs/1811.03474}{{\ttfamily arXiv:1811.03474
  [hep-ph]}}.

\bibitem{Aloni:2019ruo}
D.~Aloni, C.~Fanelli, Y.~Soreq, and M.~Williams, ``{Photoproduction of
  Axionlike Particles},''
  \href{http://dx.doi.org/10.1103/PhysRevLett.123.071801}{{\em Phys. Rev.
  Lett.} {\bfseries 123} no.~7, (2019) 071801},
  \href{http://arxiv.org/abs/1903.03586}{{\ttfamily arXiv:1903.03586
  [hep-ph]}}.

\bibitem{Bauer:2020jbp}
M.~Bauer, M.~Neubert, S.~Renner, M.~Schnubel, and A.~Thamm, ``{The Low-Energy
  Effective Theory of Axions and ALPs},''
  \href{http://dx.doi.org/10.1007/JHEP04(2021)063}{{\em JHEP} {\bfseries 04}
  (2021) 063}, \href{http://arxiv.org/abs/2012.12272}{{\ttfamily
  arXiv:2012.12272 [hep-ph]}}.

\bibitem{Bauer:2021wjo}
M.~Bauer, M.~Neubert, S.~Renner, M.~Schnubel, and A.~Thamm, ``{Consistent
  treatment of axions in the weak chiral Lagrangian},''
  \href{http://arxiv.org/abs/2102.13112}{{\ttfamily arXiv:2102.13112
  [hep-ph]}}.

\bibitem{Sakaki:2020mqb}
Y.~Sakaki and D.~Ueda, ``{Searching for new light particles at the
  international linear collider main beam dump},''
  \href{http://dx.doi.org/10.1103/PhysRevD.103.035024}{{\em Phys. Rev. D}
  {\bfseries 103} no.~3, (2021) 035024},
  \href{http://arxiv.org/abs/2009.13790}{{\ttfamily arXiv:2009.13790
  [hep-ph]}}.

\bibitem{Florez:2021zoo}
A.~Fl\'orez, A.~Gurrola, W.~Johns, P.~Sheldon, E.~Sheridan, K.~Sinha, and
  B.~Soubasis, ``{Probing axion-like particles with $\gamma \gamma$ final
  states from vector boson fusion processes at the LHC},''
  \href{http://arxiv.org/abs/2101.11119}{{\ttfamily arXiv:2101.11119
  [hep-ph]}}.

\bibitem{Brdar:2020dpr}
V.~Brdar, B.~Dutta, W.~Jang, D.~Kim, I.~M. Shoemaker, Z.~Tabrizi, A.~Thompson,
  and J.~Yu, ``{Axion-like Particles at Future Neutrino Experiments: Closing
  the ''Cosmological Triangle''},''
  \href{http://dx.doi.org/10.1103/PhysRevLett.126.201801}{{\em Phys. Rev.
  Lett.} {\bfseries 126} no.~20, (2021) 201801},
  \href{http://arxiv.org/abs/2011.07054}{{\ttfamily arXiv:2011.07054
  [hep-ph]}}.

\bibitem{Bjorken:2009mm}
J.~D. Bjorken, R.~Essig, P.~Schuster, and N.~Toro, ``{New Fixed-Target
  Experiments to Search for Dark Gauge Forces},''
  \href{http://dx.doi.org/10.1103/PhysRevD.80.075018}{{\em Phys. Rev. D}
  {\bfseries 80} (2009) 075018},
  \href{http://arxiv.org/abs/0906.0580}{{\ttfamily arXiv:0906.0580 [hep-ph]}}.

\bibitem{Donnelly:2017aaa}
T.~W. Donnelly, J.~A. Formaggio, B.~R. Holstein, R.~G. Milner, and B.~Surrow,
  {\em {Foundations of Nuclear and Particle Physics}}.
\newblock Cambridge University Press, 4, 2017.

\bibitem{Zyla:2020zbs}
{ Particle Data Group} Collaboration, P.~A. Zyla {\em et~al.}, ``{Review of
  Particle Physics},'' \href{http://dx.doi.org/10.1093/ptep/ptaa104}{{\em PTEP}
  {\bfseries 2020} no.~8, (2020) 083C01}.

\bibitem{Tsai:1973py}
Y.-S. Tsai, ``{Pair Production and Bremsstrahlung of Charged Leptons},''
  \href{http://dx.doi.org/10.1103/RevModPhys.46.815}{{\em Rev. Mod. Phys.}
  {\bfseries 46} (1974) 815}. [Erratum: Rev.Mod.Phys. 49, 421--423 (1977)].

\bibitem{Abbiendi:2002je}
{ OPAL} Collaboration, G.~Abbiendi {\em et~al.}, ``{Multiphoton production in
  e+ e- collisions at s**(1/2) = 181-GeV to 209-GeV},''
  \href{http://dx.doi.org/10.1140/epjc/s2002-01074-5}{{\em Eur. Phys. J. C}
  {\bfseries 26} (2003) 331--344},
  \href{http://arxiv.org/abs/hep-ex/0210016}{{\ttfamily arXiv:hep-ex/0210016}}.

\bibitem{Larin:2010kq}
{ PrimEx} Collaboration, I.~Larin {\em et~al.}, ``{A New Measurement of the
  $\pi^0$ Radiative Decay Width},''
  \href{http://dx.doi.org/10.1103/PhysRevLett.106.162303}{{\em Phys. Rev.
  Lett.} {\bfseries 106} (2011) 162303},
  \href{http://arxiv.org/abs/1009.1681}{{\ttfamily arXiv:1009.1681 [nucl-ex]}}.

\bibitem{Dusaev:2020gxi}
R.~R. Dusaev, D.~V. Kirpichnikov, and M.~M. Kirsanov, ``{Photoproduction of
  axionlike particles in the NA64 experiment},''
  \href{http://dx.doi.org/10.1103/PhysRevD.102.055018}{{\em Phys. Rev. D}
  {\bfseries 102} no.~5, (2020) 055018},
  \href{http://arxiv.org/abs/2004.04469}{{\ttfamily arXiv:2004.04469
  [hep-ph]}}.

\bibitem{Banerjee:2020fue}
{ NA64} Collaboration, D.~Banerjee {\em et~al.}, ``{Search for Axionlike and
  Scalar Particles with the NA64 Experiment},''
  \href{http://dx.doi.org/10.1103/PhysRevLett.125.081801}{{\em Phys. Rev.
  Lett.} {\bfseries 125} no.~8, (2020) 081801},
  \href{http://arxiv.org/abs/2005.02710}{{\ttfamily arXiv:2005.02710
  [hep-ex]}}.

\bibitem{BelleII:2020fag}
{ Belle-II} Collaboration, F.~Abudin\'en {\em et~al.}, ``{Search for Axion-Like
  Particles produced in $e^+e^-$ collisions at Belle II},''
  \href{http://dx.doi.org/10.1103/PhysRevLett.125.161806}{{\em Phys. Rev.
  Lett.} {\bfseries 125} no.~16, (2020) 161806},
  \href{http://arxiv.org/abs/2007.13071}{{\ttfamily arXiv:2007.13071
  [hep-ex]}}.

\bibitem{Aubert:2008as}
{ BaBar} Collaboration, B.~Aubert {\em et~al.}, ``{Search for Invisible Decays
  of a Light Scalar in Radiative Transitions $\upsilon_{3S} \to \gamma$ A0},''
  in {\em {34th International Conference on High Energy Physics}}.
\newblock 7, 2008.
\newblock \href{http://arxiv.org/abs/0808.0017}{{\ttfamily arXiv:0808.0017
  [hep-ex]}}.

\bibitem{Bjorken:1988as}
J.~D. Bjorken, S.~Ecklund, W.~R. Nelson, A.~Abashian, C.~Church, B.~Lu, L.~W.
  Mo, T.~A. Nunamaker, and P.~Rassmann, ``{Search for Neutral Metastable
  Penetrating Particles Produced in the SLAC Beam Dump},''
  \href{http://dx.doi.org/10.1103/PhysRevD.38.3375}{{\em Phys. Rev. D}
  {\bfseries 38} (1988) 3375}.

\bibitem{Blumlein:1990ay}
J.~Blumlein {\em et~al.}, ``{Limits on neutral light scalar and pseudoscalar
  particles in a proton beam dump experiment},''
  \href{http://dx.doi.org/10.1007/BF01548556}{{\em Z. Phys. C} {\bfseries 51}
  (1991) 341--350}.

\bibitem{Blumlein:1991xh}
J.~Blumlein {\em et~al.}, ``{Limits on the mass of light (pseudo)scalar
  particles from Bethe-Heitler e+ e- and mu+ mu- pair production in a proton -
  iron beam dump experiment},''
  \href{http://dx.doi.org/10.1142/S0217751X9200171X}{{\em Int. J. Mod. Phys. A}
  {\bfseries 7} (1992) 3835--3850}.

\bibitem{Carenza:2020zil}
P.~Carenza, O.~Straniero, B.~D\"obrich, M.~Giannotti, G.~Lucente, and
  A.~Mirizzi, ``{Constraints on the coupling with photons of heavy
  axion-like-particles from Globular Clusters},''
  \href{http://dx.doi.org/10.1016/j.physletb.2020.135709}{{\em Phys. Lett. B}
  {\bfseries 809} (2020) 135709},
  \href{http://arxiv.org/abs/2004.08399}{{\ttfamily arXiv:2004.08399
  [hep-ph]}}.

\bibitem{Raffelt:1990yz}
G.~G. Raffelt, ``{Astrophysical methods to constrain axions and other novel
  particle phenomena},''
  \href{http://dx.doi.org/10.1016/0370-1573(90)90054-6}{{\em Phys. Rept.}
  {\bfseries 198} (1990) 1--113}.

\bibitem{Bar:2019ifz}
N.~Bar, K.~Blum, and G.~D'Amico, ``{Is there a supernova bound on axions?},''
  \href{http://dx.doi.org/10.1103/PhysRevD.101.123025}{{\em Phys. Rev. D}
  {\bfseries 101} no.~12, (2020) 123025},
  \href{http://arxiv.org/abs/1907.05020}{{\ttfamily arXiv:1907.05020
  [hep-ph]}}.

\bibitem{Raffelt:1987im}
G.~Raffelt and L.~Stodolsky, ``{Mixing of the Photon with Low Mass
  Particles},'' \href{http://dx.doi.org/10.1103/PhysRevD.37.1237}{{\em Phys.
  Rev. D} {\bfseries 37} (1988) 1237}.

\bibitem{Zavattini:2007ee}
{ PVLAS} Collaboration, E.~Zavattini {\em et~al.}, ``{New PVLAS results and
  limits on magnetically induced optical rotation and ellipticity in vacuum},''
  \href{http://dx.doi.org/10.1103/PhysRevD.77.032006}{{\em Phys. Rev. D}
  {\bfseries 77} (2008) 032006},
  \href{http://arxiv.org/abs/0706.3419}{{\ttfamily arXiv:0706.3419 [hep-ex]}}.

\bibitem{Ballou:2015cka}
{ OSQAR} Collaboration, R.~Ballou {\em et~al.}, ``{New exclusion limits on
  scalar and pseudoscalar axionlike particles from light shining through a
  wall},'' \href{http://dx.doi.org/10.1103/PhysRevD.92.092002}{{\em Phys. Rev.
  D} {\bfseries 92} no.~9, (2015) 092002},
  \href{http://arxiv.org/abs/1506.08082}{{\ttfamily arXiv:1506.08082
  [hep-ex]}}.

\bibitem{Ehret:2010mh}
K.~Ehret {\em et~al.}, ``{New ALPS Results on Hidden-Sector Lightweights},''
  \href{http://dx.doi.org/10.1016/j.physletb.2010.04.066}{{\em Phys. Lett. B}
  {\bfseries 689} (2010) 149--155},
  \href{http://arxiv.org/abs/1004.1313}{{\ttfamily arXiv:1004.1313 [hep-ex]}}.

\bibitem{Andriamonje:2007ew}
{ CAST} Collaboration, S.~Andriamonje {\em et~al.}, ``{An Improved limit on the
  axion-photon coupling from the CAST experiment},''
  \href{http://dx.doi.org/10.1088/1475-7516/2007/04/010}{{\em JCAP} {\bfseries
  04} (2007) 010}, \href{http://arxiv.org/abs/hep-ex/0702006}{{\ttfamily
  arXiv:hep-ex/0702006}}.

\bibitem{Jones:2014aoa}
A.~B. Jones and B.~A. Brown, ``{Two-parameter Fermi function fits to
  experimental charge and point-proton densities for Pb208},''
  \href{http://dx.doi.org/10.1103/PhysRevC.90.067304}{{\em Phys. Rev. C}
  {\bfseries 90} no.~6, (2014) 067304}.

\bibitem{vonWeizsacker:1934nji}
C.~F. von Weizsacker, ``{Radiation emitted in collisions of very fast
  electrons},'' \href{http://dx.doi.org/10.1007/BF01333110}{{\em Z. Phys.}
  {\bfseries 88} (1934) 612--625}.

\bibitem{Williams:1934ad}
E.~J. Williams, ``{Nature of the high-energy particles of penetrating radiation
  and status of ionization and radiation formulae},''
  \href{http://dx.doi.org/10.1103/PhysRev.45.729}{{\em Phys. Rev.} {\bfseries
  45} (1934) 729--730}.

\end{thebibliography}\endgroup

\end{document}